\documentclass[twocolumn,aps,prb,final,amsfonts,amssymb,
amsmath,floatfix,showpacs,superscriptaddress]{revtex4}

\usepackage[]{graphicx}
\usepackage{bm}

\newcommand{\etal}{\textit{et al.}}
\newcommand{\eg}{e.g., }

\begin{document}

\title{Millisecond-range electron spin memory in singly-charged InP 
quantum dots}
\author{Bipul~Pal}
\email[E-mail: ]{bipulpal@sakura.cc.tsukuba.ac.jp}
\affiliation{Institute of Physics, University of Tsukuba, Tsukuba 305-8571, 
Japan}
\author{Michio~Ikezawa}
\affiliation{Institute of Physics, University of Tsukuba, Tsukuba 305-8571, 
Japan}
\author{Yasuaki~Masumoto}
\affiliation{Institute of Physics, University of Tsukuba, Tsukuba 305-8571, 
Japan}
\author{Ivan~V.~Ignatiev}
\affiliation{Institute of Physics, University of Tsukuba, Tsukuba 305-8571, 
Japan}
\affiliation{Institute of Physics, St. Petersburg State University, 
St.-Petersburg 198504, Russia}   
\date{\today}

\begin{abstract}
We report millisecond-range spin memory of resident electrons in an 
ensemble of InP quantum dots (QDs) under a small magnetic field of $0.1$~T 
applied along the optical excitation axis at temperatures up to about 5~K. 
A pump-probe photoluminescence (PL) technique is used for optical 
orientation of electron spins by the pump pulses and for study of spin 
relaxation over the long time scale by measuring the degree of circular 
polarization of the probe PL as a function of pump-probe delay. Dependence 
of spin decay rate on magnetic field and temperature suggests two-phonon 
processes as the dominant spin relaxation mechanism in this QDs at low 
temperatures. 
\end{abstract}


\maketitle

Electron spin in semiconductors may be suitable for use as a quantum memory 
in quantum repeaters as semiconductors are capable of converting photons to 
electrons (and holes) while transferring quantum information from photon 
polarization to electron spin, and vice versa.~\cite{awschalombook} However, 
a long spin relaxation time ($\tau_{s}$) is necessary to realize spin 
quantum memory. In semiconductor quantum dots (QDs) $\tau_{s}$ up to a few 
ms was theoretically predicted as a result of suppression of important spin 
relaxation mechanisms due to 3D confinement and lack of energy 
dispersion.~\cite{khaetskiiprb61,khaetskiiprb64,woodsprb66} This has 
stimulated experimental investigations of long spin memory in 
QDs.~\cite{cortezprl89,fujisawanature419,elzermannature430,%
kroutvarnature432,laurentphe20,coltonprb69,ikezawaprbsub} 
For example, electron spin relaxation time of about 200~$\mu$s in InGaAs 
vertical QDs~\cite{fujisawanature419} at $T \le 0.5$~K and zero external 
magnetic field ($B$), about 0.85~ms in electrically gated GaAs lateral 
QDs~\cite{elzermannature430} at $T<0.3$~K and $B=8$~T, and about 20~ms 
in self-assembled InGaAs QDs~\cite{kroutvarnature432} at $T=1$~K and 
$B=4$~T has been recently reported. Many of the previous results of long 
spin memory in QDs were obtained in InGaAs and GaAs QDs under special 
condition of low temperature ($\le 1$~K) and large  magnetic field 
($\ge 4$~T). 

\begin{figure}[htb]
\includegraphics[clip,width=6.5cm]{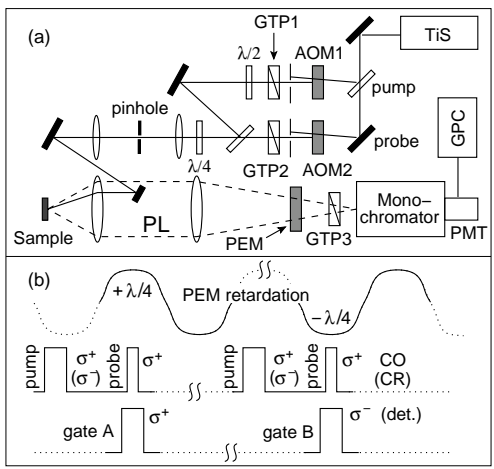}
\caption{\label{setup} Schematic of the experimental setup (a) and time 
synchronization (b) of the PEM retardation, the probe pulses, and the 
GPC gates. 
}
\end{figure}

In this letter we report observation of optically created electron 
spin-orientation surviving up to about 1~ms in an ensemble of singly 
negatively charged InP QDs at $B=0.1$~T applied along the optical 
excitation axis at $T \sim 5$~K. The sample consists of a single layer 
of self-assembled InP QDs embedded between GaInP barriers. The average 
base diameter (height) of the QDs is about 40~(5)~nm with an areal density 
$\sim 10^{10}$~cm$^{-2}$. We use a pump-probe PL 
technique~\cite{cortezprl89,coltonprb69,ikezawaprbsub} to study electron 
spin orientation dynamics by measuring the circular polarization [defined 
as $P=(I^{++}-I^{+-})/(I^{++}+I^{+-})$, where $I^{++(-)}$ is the PL 
intensity for excitation with $\sigma^{+}$ probe and detection of 
$\sigma^{+(-)}$ probe PL] of the probe pulse PL in presence of a 
preexcitation by a pump pulse. Our experimental setup is schematically 
shown in Fig.~\ref{setup}(a).

A CW Ti:sapphire laser beam is split into pump and probe beams. Two 
acousto-optic modulators (AOM) driven by programmable function generator 
(PFG) generates pump and probe pulses with controllable 
pulse width and delay ($\tau$) between them. 
Glan-Thompson polarizers (GTP) and wave plates are used to control the 
circular polarization of the pump and probe beams. 
The PL signal is sent through a combination of a photo-elastic 
modulator (PEM) and a GTP before dispersing in a monochromator and 
detecting in a GaAs photomultiplier tube (PMT). The PMT output is connected 
to a two-channel gated photon counter (GPC). The PEM acts as an oscillating 
$\lambda/4$-plate and when combined with GTP, allows detection of PL 
intensity in the $\sigma^{+}$ and $\sigma^{-}$ channels. The PEM frequency, 
$f_{\text{P}}=42$~kHz, is reduced to $f_{\text{T}}=f_{\text{P}}/(n+0.5)$ 
(typically $n=40$ in our measurements) to trigger the PFG and 
GPC.~\cite{kalevich} Thus, one probe pulse (and A-gate of the GPC) is 
centered at the $+\lambda/4$ and the next probe pulse (and B-gate of the 
GPC) is centered at the $-\lambda/4$ retardation peaks of the PEM 
[Fig.~\ref{setup}(b)]. We typically use a GPC gate width of 5~$\mu$s, 
while the pump (probe) pulse width is 60~(3)~$\mu$s, giving a pump (probe) 
power density [$W_{\text{pump (probe)}}$] of about 0.5~(0.05)~W~cm$^{-2}$. 
The low probe power density ensures that the pump-induced spin polarization 
is not fully erased by the probe pulse. 
The excitation energy is tuned to about 1.771~eV (below-barrier, QD excited 
state excitation) and the QD ground state PL is detected at about 1.729~eV. 
An external electric bias of $U_{\text{bias}}=-0.1$~V is applied to the sample. 
We find that under this condition the PL polarization is negative and reaches 
maximum.~\cite{ikezawaprbsub} A study of trionic quantum beats in this 
sample~\cite{kozinprb65} showed that at $U_{\text{bias}} \approx -0.1$~V 
each QD contains one resident electron on an average. This suggests that 
the negative PL polarization arises from trionic state, as is discussed \eg 
in Refs.~\onlinecite{cortezprl89,laurentphe20,kavokinpssa195,brackerprl94}. 

\begin{figure}[htb]
\includegraphics[clip,width=8.3cm]{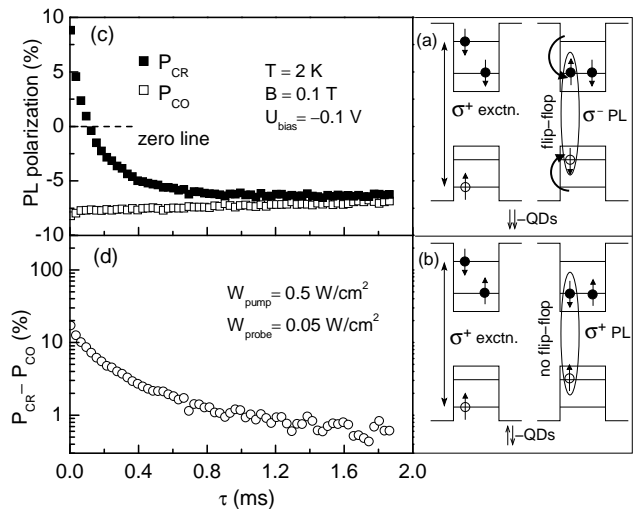}
\caption{\label{longspin} Schematics of $\downarrow \! \downarrow$-QDs 
(a) and $\uparrow \! \downarrow$-QDs (b). Probe 
PL polarization for co- ($P_{\text{CO}}$) and cross- ($P_{\text{CR}}$) 
circularly polarized pump-probe (c), and the difference 
\mbox{$P_{\text{CR}}-P_{\text{CO}}$} (d) as a function of $\tau$. 
}
\end{figure}

In our experiments, a $\sigma^{+}$ (or $\sigma^{-}$) polarized pump 
induces $\downarrow$ (or $\uparrow$) spin orientation of the resident 
electrons.~\cite{opticalorientation} 
A probe pulse, variably delayed with respect to the 
pump pulse, tests this pump-induced spin-orientation. The probe beam (always 
$\sigma^{+}$ polarized) creates a hot trion with \emph{parallel} 
[$\downarrow \! \downarrow$-QDs] or \emph{anti-parallel} 
[$\uparrow \! \downarrow$-QDs] electron spins. After a flip-flop process in 
$\downarrow \! \downarrow$-QDs shown schematically in Fig.~\ref{longspin}(a), 
the probe PL polarization becomes negative,~\cite{ikezawaprbsub} while it 
is positive for the $\uparrow \! \downarrow$-QDs [Fig.~\ref{longspin}(b)]. 
At any given $\tau$ the net probe PL polarization is determined by the ratio 
of $\downarrow \! \downarrow$- and $\uparrow \! \downarrow$-QDs. 

We measure the probe PL polarization for (i) co-circularly polarized 
pump-probe ($P_{\text{CO}}$) [pump  creates more 
$\downarrow \! \downarrow$-QDs] and (ii) cross-circularly polarized 
pump-probe ($P_{\text{CR}}$) [pump creates more 
$\uparrow \! \downarrow$-QDs] [pump and probe polarizations for the two 
cases are indicated in Fig.~\ref{setup}(b)]. 
A small static magnetic field of $B=0.1$~T is applied parallel to the 
optical excitation (and sample growth) axis to suppress the effect of 
fluctuating nuclear magnetic field.~\cite{merkulovprb65,braunprl94}  
Polarizations $P_{\text{CR}}$ and $P_{\text{CO}}$ as a function of 
$\tau$ are shown in Fig.~\ref{longspin}(c). 
The difference \mbox{$P_{\text{CR}}-P_{\text{CO}}$} is a good measure of 
the pump induced spin orientation of the resident electrons.~\cite{foot1} 
A semilogarithmic plot of \mbox{$P_{\text{CR}}-P_{\text{CO}}$} obtained 
from Fig.~\ref{longspin}(c) shows that the spin memory decay is 
nonexponential [Fig.~\ref{longspin}(d)]. Thus, a spin relaxation time 
cannot be defined in a simple way. However, it is clear from this data 
that the spin memory decays on a millisecond time-scale.  

The observed long-lived spin polarization could result from a dynamic 
nuclear polarization which may appear under the experimental condition 
used.~\cite{merkulovprb65,optorient,gammonprl86} However, in a recent 
study~\cite{ikezawaprbsub} of this aspect we have shown that very small 
effective magnetic field ($<0.02$~T) in InP QDs,~\cite{ignatievconf,%
dzhioevpss41} arising from dynamic nuclear polarization, is not 
consistent with the large amplitude of PL polarization observed in this 
sample. Thus, the long spin memory observed here should be related to 
the lack of efficient spin decay path in QDs. To investigate the spin 
relaxation mechanisms effective in this case we study temperature 
and magnetic field dependence of the spin decay process. 

\begin{figure}[htb]
\includegraphics[clip,width=6.6cm]{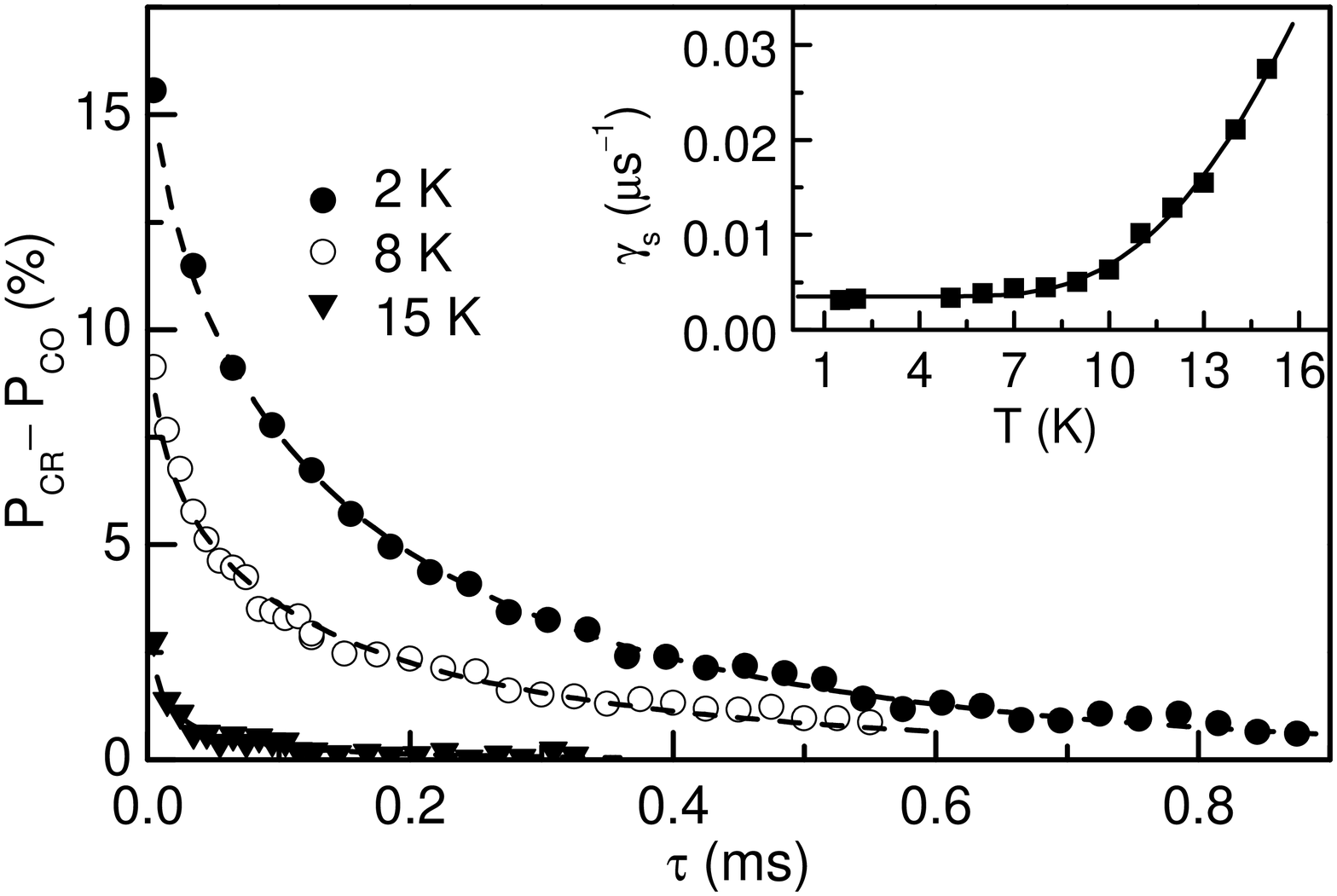}
\caption{\label{tdep} Delay dependence of 
\mbox{$P_{\text{CR}}-P_{\text{CO}}$} at a few $T$. $B=0.1$~T, 
$U_{\text{bias}}=-0.1$~V, $W_{\text{pump (probe)}}=0.5$ (0.05)~W~cm$^{-2}$. 
Dashed lines are stretched exponential fits (discussed in the text). Inset 
shows spin decay rate ($\gamma_{s}$) as a function of $T$.
}
\end{figure}

Figure~\ref{tdep} shows decay of \mbox{$P_{\text{CR}}-P_{\text{CO}}$} 
at a few temperatures for $B=0.1$~T. A faster decay is seen with 
increasing $T$. As noted earlier, the decay is nonexponential and  
suggests a distribution of decay rates, which may arise due to 
inhomogeneous environment and size-distribution of the 
QDs.~\cite{ikezawaprbsub} Theoretical 
analysis shows (see \eg Ref.~\onlinecite{chenlum102}) 
that a spread of the relaxation rate results in a nonexponential decay 
of the form $\sim \exp[-(\gamma_{s} \tau)^{c}]$ (the so-called stretched 
exponential function), where the parameter $c$  depends on the 
physical processes causing the spread. We find that the function fits 
our data very well (dashed lines in Fig.~\ref{tdep}) if we use $c$ as a 
fitting parameter. Effective spin decay rate $\gamma_{s}$ obtained from 
such fits is plotted in the inset of Fig.~\ref{tdep} as a function of $T$. 
A rapid increase in $\gamma_{s}$ is seen for $T>8$~K. Such an increase is 
expected for thermally activated spin relaxation due to the phonon-mediated 
coupling of the ground and excited electron states (two-phonon Orbach 
process).~\cite{abragambook} We find that the function 
$\gamma_{s} \sim (\exp[\Delta E/k_{B}T]-1)^{-1}+\gamma_{0}$ 
($\Delta E =$~activation energy, $k_{B} =$~Boltzmann constant, and 
$\gamma_{0}$ stands for spin decay rate 
arising from temperature independent relaxation mechanisms) describing this 
process fits the data very well (solid line in the inset of Fig.~\ref{tdep}). 
From the fit we obtain $\Delta E \approx 5$~meV. This value is smaller than 
that obtained experimentally for electron level spacing of 15 meV in 
Ref.~\onlinecite{kozinprb65}. This discrepancy is probably due to 
the difference in QD sub-ensemble probed in the two cases. 

\begin{figure}[htb]
\includegraphics[clip,width=6.6cm]{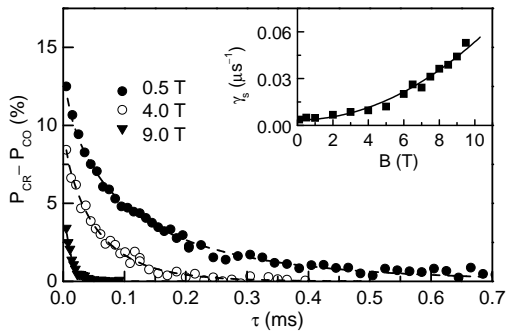}
\caption{\label{bdep} Delay dependence of 
\mbox{$P_{\text{CR}}-P_{\text{CO}}$} at a few $B$. $T=2$~K, 
$U_{\text{bias}}=-0.1$~V, $W_{\text{pump (probe)}}=0.5$ (0.05)~W~cm$^{-2}$. 
Dashed lines are stretched exponential fits (discussed in the text). Inset 
shows spin decay rate ($\gamma_{s}$) as a function of $B$.
}
\end{figure}

We now present the magnetic field dependence of the spin decay 
process. Decay of \mbox{$P_{\text{CR}}-P_{\text{CO}}$} at 
a few values of $B$ at $T=2$~K is shown in Fig.~\ref{bdep}. We find 
that the decay becomes increasingly faster with increase in $B$. An 
effective spin decay rate obtained from stretched exponential fit to 
the data is plotted as a function of $B$ in the inset of Fig.~\ref{bdep}. 
Decay rate $\gamma_{s}$ is found to increase superlinearly with $B$. 
Several possible mechanisms for such an increase are discussed in the 
literature.~\cite{khaetskiiprb64,woodsprb66,erlingssonprb66} Magnetic 
filed couples the higher energy states with nonzero orbital momentum to 
the electron spin states split by the magnetic field (Zeeman splitting) 
that allows a small admixture of the states of opposite spin to each Zeeman
sublevel.~\cite{khaetskiiprb64,woodsprb66} At low temperature this 
enables spin-flip transition between Zeeman sublevels via participation 
of acoustic phonons to dissipate energy (one-phonon resonant process). 
With increasing $B$ the Zeeman splitting increases. Due 
to higher density of resonant phonons at increased energy and more 
efficient mixture of the states by the magnetic field, the spin relaxation 
rate increases. Theoretical calculations~\cite{khaetskiiprb64,woodsprb66} 
have predicted $\gamma_{s} \sim B^{5}$ at very low temperature and large 
magnetic field if the spin-orbit interaction and the one-phonon scattering 
dominate. However, for $T$ of about a few kelvin, the two-phonon nonresonant 
(Raman) scattering may become important.~\cite{woodsprb66} In that case, 
the magnetic field dependence is only determined by the admixture of the 
excited states and becomes quadratic.~\cite{khaetskiiprb64} Our data in 
Fig.~\ref{bdep}-inset can be fitted very well with 
$\gamma_{s}=\alpha +\beta B^{2}$. This argues for the two-phonon scattering 
as the main mechanism of acceleration of the spin relaxation in magnetic 
field. 

The acceleration of spin relaxation could also result from hyperfine 
interaction.~\cite{erlingssonprb66} However, this is unlikely in our case 
due to very small nuclear spin polarization in the InP 
QDs~\cite{ignatievconf,dzhioevpss41} we studied.  

In conclusion, we have observed long spin memory, persisting over 1~ms, 
in an ensemble of singly negatively charged InP QDs at small magnetic 
field (0.1~T) and at moderate temperature ($\sim 5$~K). Our data on the 
magnetic field and temperature dependence of spin decay rate suggests 
two-phonon scattering may be the dominant spin relaxation mechanism. 
Long spin memory observed here in III-V semiconductor QDs is relevant 
for quantum information communication and storage. Though our study is 
made at about 0.7~$\mu$m wavelength ($\lambda$), III-V semiconductor 
system can be easily adapted to $\lambda=1.3$ and 1.5~$\mu$m, suitable 
for fiber optic communication. 

Authors thank I.~Ya.~Gerlovin and T.~Takagahara for fruitful discussions. 
B.~Pal thanks INOUE Foundation for Science, Japan, for financial 
support. This work is supported by the Grant-in-Aid for the Scientific 
Research \#13852003 and \#16031203 from the MEXT, Japan.


\begin{thebibliography}{99}
\bibitem{opticalorientation} Various mechanisms of optical orientation of 
spins are discussed \eg in Refs.~\protect{\onlinecite{cortezprl89,%
brackerprl94,optorient}}. Possible mechanisms of optical orientation of 
resident electron spins in InP QDs under study will be discussed elsewhere.
\bibitem{foot1} Effect of dynamic nuclear polarization (if present) is 
expected to be the same for $P_{\text{CR}}$ and $P_{\text{CO}}$ and taking 
their difference may eliminate the influence of dynamic nuclear 
polarization in the range of our measurement. A direct relation between 
\mbox{$P_{\text{CR}}-P_{\text{CO}}$} and the spin-orientation of resident 
electrons will be discussed elsewhere.
\end{thebibliography}
\end{document}